\documentclass[11pt,showpacs,preprintnumbers,amsmath,amssymb,prd,nofootinbib,superscriptaddress]{revtex4-2}

\usepackage{dcolumn}
\usepackage{bm}
\usepackage{ifpdf}
\usepackage{hyperref}
\usepackage{bm}
\usepackage{xcolor,color,graphicx,graphics}
\usepackage[spanish,english]{babel}
\usepackage[latin1]{inputenc}
\usepackage[OT1]{fontenc}
\usepackage{latexsym,amssymb,amsmath,amsfonts}
\usepackage{makeidx}
\usepackage{epsfig,subfigure}
\usepackage{natbib}
\usepackage{epstopdf}
\usepackage{mathrsfs}
\usepackage{hyperref}
\hypersetup{colorlinks=true, linkcolor=blue, citecolor=green}
\usepackage{enumerate}

\everymath{\displaystyle}
\usepackage{graphicx}

\usepackage[T1]{fontenc}
\usepackage{amsmath}
\usepackage{amssymb}
\usepackage{graphicx}
\usepackage{xcolor}

\newcommand{\bea}{\begin{eqnarray}}
\newcommand{\eea}{\end{eqnarray}}

\newcommand{\orcid}[1]{\href{https://orcid.org/#1}{\includegraphics[width=10pt]{orcid}}}

\usepackage{fixmath}

\begin{document}

\title{TFD formalism: applications to the scalar field in a Lorentz-violating theory}

\author{L. H. A. R. Ferreira}
\email{luiz.ferreira@fisica.ufmt.br }
\affiliation{Instituto de F\'{\i}sica, Universidade Federal de Mato Grosso,\\
78060-900, Cuiab\'{a}, Mato Grosso, Brazil}

\author{A. F. Santos} 
\email{alesandroferreira@fisica.ufmt.br}
\affiliation{Instituto de F\'{\i}sica, Universidade Federal de Mato Grosso,\\
78060-900, Cuiab\'{a}, Mato Grosso, Brazil}

\author{Faqir C. Khanna \footnote{Professor Emeritus - Physics Department, Theoretical Physics Institute, University of Alberta\\
Edmonton, Alberta, Canada}}
\email{fkhanna@ualberta.ca; khannaf@uvic.ca}
\affiliation{Department of Physics and Astronomy, University of Victoria,\\
3800 Finnerty Road, Victoria BC V8P 5C2, Canada}

\begin{abstract}

The Thermofield Dynamics (TFD) formalism is considered. In this context, a Lorentz-breaking scalar field theory is introduced. In contrast to the Matsubara formalism, the best-known approach to introducing the temperature effect, TFD is a real-time formalism and is a topological field theory. While in Matsubara the temperature effects are introduced as a consequence of a compactification of the field in a finite interval on the time axis, in the TFD this effect emerges through a condensed state related to the Bogoliubov transformation. An advantage of the TFD formalism is that different topologies, which lead to different effects, can be chosen. Here, three different topologies are considered. Then the Stefan-Boltzmann law and Casimir effect at zero and non-zero temperature are calculated. This is a unique feature of TFD, which allows us to treat different phenomena in the same way.


\end{abstract}

\maketitle

\section{Introduction}

Temperature is a natural ingredient in all phenomena described by nature. However, to study the effects of temperature, a thermal theory is needed. One of the best-known theories for introducing temperature is the Matsubara formalism, also known as the imaginary time formalism \cite{Matsubara}. This approach implies a replacement of the coordinate of time by imaginary time. Another thermal formalism is the real-time formalism which allows to analyze the temporal evolution of the system and temperature effects together. It is divided into two approaches: (i) the closed-time path formulation which consists of the doubling of degrees of freedom that modifies the Green functions \cite{Schwinger}, and (ii) the Thermofield Dynamics (TFD) formalism proposed by Takahashi and Umezawa \cite{Umezawa1, Umezawa2, Umezawa22, Khanna0, Khanna1, Khanna2, GBT}. In this approach, a thermal vacuum state is constructed. This implies that the vacuum expectation value of an operator is equal to its statistical average. Here, the TFD formalism is used.

To build the TFD formalism two fundamental elements are required: the duplication of the Hilbert space and the Bogoliubov transformation. The expanded Hilbert space is composed of the standard Hilbert space ${\cal S}$ and the tilde space $\tilde{\cal S}$, i.e. ${\cal S}_T={\cal S}\otimes \tilde{\cal S}$. These two spaces are related through the tilde conjugation rules. Then an operator in the original Hilbert space is associated with two operators in the thermal or double Hilbert space. The other key ingredient is the Bogoliubov transformation, which is a rotation that involves the tilde and non-tilde operators. From this, the effects of temperature are introduced. Furthermore, this formalism is a topological field theory described by $\Gamma_D^d=(\mathbb{S}^1)^d\times \mathbb{R}^{D-d}$, where $D$ are the space-time dimensions and $d$ is the number of compactified dimensions. This implies that any number of dimensions of the manifold can be compactified into the hyper-torus $(\mathbb{S}^1)^d$. Then both effects, due to thermal and spatial compactifications, can be investigated. In this paper, the scalar field is considered in three different topologies: (i) the time coordinate is compactified into a circle of length $1/T$, where $T$ is the temperature; (ii) the $z$ coordinate is compactified into a circumference of length $L$, and (iii) both coordinates, i.e. $t$ and $z$ are compactified. It is interesting to note that this formalism allows us to explore some different phenomena, such as the Stefan-Boltzmann law and the Casimir effect, in the same way. Using this theory, let us study the Casimir effect at zero and finite temperature for the scalar field in a Lorentz-violating theory.

Lorentz invariance emerges as a non-exact symmetry in models that describe a fundamental theory that arises at very high energy scales \cite{Kostelecky1, Kostelecky2}. A small violation of Lorentz symmetry leads to a new physics beyond the standard model. To study the effects due to the violation of Lorentz symmetry, an effective field theory, called Standard Model Extension (SME), has been constructed \cite{Kostelecky3, Kostelecky4, Kostelecky5}. The SME is composed of general relativity, the standard model, and all terms that violate Lorentz. Terms that belong to the SME are classified as CPT-odd and CPT-even. The CPT-odd term violates Lorentz symmetry and CPT (Charge conjugation, Parity and Time reversal) symmetry, while the CPT-even term violates Lorentz symmetry, but CPT symmetry is maintained. Here, the main interest is the CPT-even aether-like term. This term was first introduced in a five-dimensional context and its consequences for different fields have been analyzed \cite{Carroll}. Some studies with such a Lorentz-violating term have been explored. For example: stability conditions have been investigated \cite{stab}, phenomenological properties of this theory have been studied \cite{phen}, the generation of this term in three, four and five space-time dimensions has been done \cite{Albert1}, a non-abelian version in four dimensions has been constructed \cite{Albert2}, among others. In this context, the aether-like Lorentz-breaking scalar field theory is considered. Then, some corrections due to Lorentz symmetry breaking are calculated for the Casimir effect.

The Casimir effect, proposed in 1948 by H. Casimir \cite{Casimir}, is a quantum phenomenon that arises due to the vacuum fluctuations of any field that exhibits a quantum nature. This effect is a consequence of boundary conditions or topology under the quantum fields. Its experimental confirmation takes place a few years after theoretical construction \cite{Sparnaay}. Over the years its accuracy has greatly increased \cite{Lamoreaux, Mohideen}. Our main objective is to calculate the Casimir effect for the scalar field in four dimensions with the aether-like term at zero and finite temperature. There are some studies that investigate the Casimir effect for a scalar field at zero and finite temperature in a theory where the Lorentz invariance is broken due to the aether term \cite{Cruz1, Cruz2, Escobar1, Escobar2, Ruiz, our}. In each of these works, different techniques and different analyzes are used. However, these applications do not exhaust the possibilities of studying this effect using different tools. Here, the Casimir effect is calculated using the TFD formalism. Through different topologies, it is possible to analyze this effect at zero and non-zero temperature. Furthermore, the calculations are carried out in a natural covariant way. This aspect represents ease of calculation, such that the results are direct and accurate. Another advantage of the TFD is that the Stefan-Boltzmann law associated with the field can be calculated in the same way, just by choosing an appropriate topology.

This paper is organized as follows. In section II, the TFD formalism is introduced. In section III, the model that describes a massive scalar field with an aether-like term that violates the Lorentz symmetry is presented. The energy-momentum tensor associated with this theory in the TFD approach is calculated. In section IV, some finite temperature and Lorentz violation applications are analyzed. Considering a time-like and space-like constant vector, the Stefan-Boltzmann law and the Casimir effect at zero and non-zero temperatures are obtained. In section V, some concluding remarks are discussed.

\section{TFD formalism}

TFD formalism is a thermal quantum field theory that allows interpreting the statistical average of an arbitrary operator as its vacuum expectation value in a thermal ground state \cite{Umezawa1, Umezawa2, Umezawa22, Khanna0, Khanna1, Khanna2, GBT}. In addition, TFD is classified as real-time formalism. Then, the effects of temperature and temporal evolution of a system are investigated together. 

In order to build a thermal vacuum, the degrees of freedom are doubled and, as a consequence, the Fock space is extended, that is, the thermal Fock space ${\cal S}_T$ is composed of the original Fock ${\cal S}$ and the tilde or dual space  $\tilde{\cal S}$, i.e. ${\cal S}_T={\cal S}\otimes \tilde{\cal S}$. There is a mapping between the two spaces defined by the tilde conjugation rules which are given as
\bea
({\cal A}_i{\cal A}_j)^\thicksim &=& \tilde{{\cal A}_i}\tilde{{\cal A}_j},\nonumber\\(c{\cal A}_i+{\cal A}_j)^\thicksim &=& c^*\tilde{{\cal A}_i}+\tilde{{\cal A}_j}, \nonumber\\ ({\cal A}_i^\dagger)^\thicksim &=& \tilde{{\cal A}_i}^\dagger, \nonumber\\ (\tilde{{\cal A}_i})^\thicksim &=& -\xi {\cal A}_i,
\eea
with $\xi = -1$ for bosons and $\xi = +1$ for fermions and ${\cal A}$ is an arbitrary operator.

In addition to the duplication of Fock space, another tool is required in this formalism. The Bogoliubov transformation consists of a rotation, between the tilde and non-tilde operators, which introduces the dependency of a new parameter on that operator. This parameter, called the compactification parameter, is defined as $\alpha=(\alpha_0,\alpha_1,\cdots\alpha_{D-1})$. The effects of temperature are introduced by taking $\alpha_0=\beta=1/T$ and all other components equal to zero. 

As an example of applying the Bogoliubov transformation, consider two arbitrary operators ${\cal O}$ and $\tilde{\cal O}$ in ${\cal S}$ and $\tilde{\cal S}$, respectively. Then
\bea
\left( \begin{array}{cc} {\cal O}(k,\alpha)  \\\xi \tilde {\cal O}^\dagger(k,\alpha) \end{array} \right)={\cal U}(\alpha)\left( \begin{array}{cc} {\cal O}(k)  \\ \xi\tilde {\cal O}^\dagger(k) \end{array} \right),
\eea
where ${\cal U}(\alpha)$ is defined as
\bea
{\cal U}(\alpha)=\left( \begin{array}{cc} u(\alpha) & -w(\alpha) \\
\xi w(\alpha) & u(\alpha) \end{array} \right),
\eea
with $u(\alpha)$ and $v(\alpha)$ related to the Bose distribution, i.e. $v^2(\alpha)=\frac{1}{e^{\alpha\omega}-1}$ and $u^2(\alpha)=1+v^2(\alpha)$.

For the calculations that are developed in the next section, the modifications caused by the introduction of the $\alpha$ parameter in the propagator of the scalar field are investigated. In the TFD formalism, this propagator is written as
\bea
G_0^{(ab)}(x-x';\alpha)=i\langle 0,\tilde{0}| \tau[\phi^a(x;\alpha)\phi^b(x';\alpha)]| 0,\tilde{0}\rangle,
\eea
where $a,b$ $=1,2$, $\tau$ is the time ordering operator and $\phi(x;\alpha)={\cal U}(\alpha)\phi(x){\cal U}^{-1}(\alpha)$. In the thermal vacuum $|0(\alpha)\rangle={\cal U}(\alpha)|0,\tilde{0}\rangle$, the propagator is
\bea
G_0^{(ab)}(x-x';\alpha)=i\int \frac{d^4k}{(2\pi)^4}e^{-ik(x-x')}G_0^{(ab)}(k;\alpha),
\eea
with $G_0^{(ab)}(k;\alpha)={\cal U}(\alpha)G_0^{(ab)}(k){\cal U}^{-1}(\alpha).$ It is important to note that, the physical Green function is given by the non-tilde part, i.e.
\bea
G_0^{(11)}(k;\alpha)=G_0(k)+ v^2(k;\alpha)[G^*_0(k)-G_0(k)],
\eea
where $v^2(k;\alpha)$ is the generalized Bogoliubov transformation \cite{GBT} defined as
\bea
v^2(k;\alpha)&=&\sum_{s=1}^p\sum_{\lbrace\sigma_s\rbrace}2^{s-1}\sum_{l_{\sigma_1},...,l_{\sigma_s}=1}^\infty(-\xi)^{s+\sum_{r=1}^sl_{\sigma_r}}\exp\left[{-\sum_{j=1}^s\alpha_{\sigma_j} l_{\sigma_j} k^{\sigma_j}}\right],\label{BT}
\eea
with $p$ being the number of compactified dimensions, $\lbrace\sigma_s\rbrace$ denotes the set of all combinations with $s$ elements and $k$ is the 4-momentum. 

\section{The energy-momentum tensor in the TFD formalism}

Here, the main idea is to obtain the energy-momentum tensor for the scalar field within a Lorentz-violating theory. The term that leads to invariance breaking is the aether-like term. In the TFD formalism, the energy-momentum tensor is thermalized and, as a consequence, some different physical quantities are calculated at finite temperature. 

The complete Lagrangian in the TFD approach, which describes the scalar field in the presence of the aether-like term, is given as
\bea
\widehat{{\cal L}}&=&{\cal L}-\widetilde{{\cal L}}\nonumber\\
&=&\frac{1}{2}\partial_\mu\phi\partial^\mu\phi+\frac{\lambda}{2}(u\cdot\partial\phi)^2+\frac{1}{2}m^2\phi^2-\frac{1}{2}\partial_\mu\tilde{\phi}\partial^\mu\tilde{\phi}+\frac{\lambda}{2}(u\cdot\partial\tilde{\phi})^2+\frac{1}{2}m^2\tilde{\phi}^2\label{1.1},
\eea
where ${\cal L}$ and $\widetilde{\cal L}$ are the Lagrangians in the original and tilde spaces, respectively. The violation of Lorentz symmetry is introduced through the term $\lambda(u\cdot\partial\phi)^2$, with $\lambda$ being a very small dimensionless parameter which determines the effects of symmetry breaking in the theory and $u^\mu$ is a constant vector that acts as a background field. 

In order to obtain the energy-momentum tensor, only the non-tilde variables in the Lagrangian are considered. However, a similar development is performed with the tilde part. Then, using the definition
\bea
T^{\mu\nu}=\frac{\partial{\cal L}}{\partial(\partial_\mu\phi)}\partial^\nu\phi-g^{\mu\nu}{\cal L},\label{1.2}
\eea
the energy-momentum tensor associated with the Lorentz-violating scalar field is 
\bea
T^{\mu\nu}=\partial^\mu\phi\partial^\nu\phi+\lambda u^\mu\partial^\nu\phi(u\cdot\partial\phi)-\frac{1}{2}g^{\mu\nu}\partial_\rho\phi\partial^\rho\phi-\frac{\lambda}{2}g^{\mu\nu}(u\cdot\partial\phi)^2-\frac{1}{2}g^{\mu\nu}m^2\phi^2\label{1.3}.
\eea
It is important to note that the product of the field operators at the same space-time points should be avoided. To solve this problem, the energy-momentum tensor is written as 
\bea
T^{\mu\nu}(x)&=&\lim_{x'\to x}\tau{\bigg\{}\partial^\mu\phi(x)\partial'^\nu\phi(x')+\lambda u^\mu\partial^\nu\phi(x)\left(g_{\rho\epsilon}u^\epsilon\partial'^\rho\phi(x')\right)-\frac{1}{2}g^{\mu\nu}g_{\gamma\rho}\partial^\gamma\phi(x)\partial'^\rho\phi(x')\nonumber\\
&-&\frac{\lambda}{2}g^{\mu\nu}g_{\sigma\omega}g_{\lambda\theta}u^\omega u^\theta\partial^\sigma\phi(x)\partial'^\lambda\phi(x')-\frac{1}{2}g^{\mu\nu}m^2\phi(x)\phi(x'){\bigg\}}\label{1.4},
\eea
where $x'$ and $x$ represent different points of space-time. Using the commutation relation
\bea
[\phi(x),\partial'^\mu\phi^\dagger(x')]=in^\mu_0\delta({\bf x-x'})\label{1.6},
\eea
with $n^\mu_0=(1,0,0,0)$ being a time-like vector, eq. (\ref{1.4}) becomes
\bea
T^{\mu\nu}(x)=\lim_{x'\to x}{\bigg\{}\Gamma^{\mu\nu}\tau[\phi(x)\phi(x')]+I^{\mu\nu}\delta({\bf x-x'}){\bigg\}},
\eea
where 
\bea
\Gamma^{\mu\nu}&=&\partial^\mu\partial'^\nu+\lambda g_{\alpha\epsilon} u^\mu u^\epsilon\partial^\nu\partial'^\alpha-\frac{1}{2}g^{\mu\nu}g_{\gamma\rho}\partial^\gamma\partial'^\rho-\frac{\lambda}{2}g^{\mu\nu}g_{\sigma\omega}g_{\lambda\theta}u^\omega u^\theta\partial^\sigma\partial'^\lambda-\frac{1}{2}g^{\mu\nu}m^2,\label{1.8}\\
I^{\mu\nu}&=&in^\mu_0n^\nu_0+i\lambda g_{\alpha\epsilon}u^\mu u^\epsilon n^\nu_0n^\alpha_0-\frac{i}{2}g^{\mu\nu}g_{\gamma\rho}n^\gamma_0n^\rho_0-\frac{\lambda}{2}ig^{\mu\nu}g_{\sigma\omega}g_{\lambda\theta}u^\omega u^\theta n^\sigma_0n^\lambda_0\label{1.9}.
\eea

Now, let us take the vacuum expectation value of the energy-momentum tensor. Then
\bea
\bigl\langle T^{\mu\nu}(x)\bigl\rangle&\equiv&\bigl\langle 0|T^{\mu\nu}|0\bigl\rangle\nonumber\\
&=&\lim_{x'\to x}\bigg\{\Gamma^{\mu\nu}\bigl\langle0|\tau[\phi(x)\phi(x')]|0\bigl\rangle+I^{\mu\nu}\delta({\bf x-x'})\bigl\langle0|0\bigl\rangle\bigg\}\label{1.11},
\eea
with
\bea
\bigl\langle0|\tau[\phi(x)\phi(x')]|0\bigl\rangle&=&-G_0(x-x')\nonumber\\
&=&-\frac{im}{4\pi^2}\frac{K_1(m\sqrt{-(x-x')^2})}{\sqrt{-(x-x')^2}}\label{1.13}
\eea
being the definition of the massive scalar field propagator. Here $K_n(x)$ is the modified Bessel function.

Introducing the TFD formalism notation as presented in the previous section, eq. (\ref{1.11}) reads
\bea
\bigl\langle T^{\mu\nu(ab)}(x;\alpha)\bigl\rangle&=&\lim_{x'\to x}\bigg\{\Gamma^{\mu\nu}G_0^{ab}(x-x';\alpha)+I^{\mu\nu}\delta({\bf x-x'})\bigg\}.
\eea

There is an important observation that must be made in the vacuum expectation value. This quantity is divergent for both cases, with or without the $\alpha$ parameter. Then a renormalization procedure is required. This prescription consists of
\bea
{\cal T^{}}^{\mu\nu(ab)}(x;\alpha)&\equiv&\bigl\langle T^{\mu\nu(ab)}(x;\alpha)\bigl\rangle-\bigl\langle T^{\mu\nu(ab)}(x)\bigl\rangle\nonumber\\
&=&\lim_{x'\to x}{\bigg\{}i\Gamma^{\mu\nu}\overline{G}^{(ab)}_0(x-x';\alpha){\bigg\}}\label{1.15},
\eea
where
\bea
\overline{G}^{(ab)}_0(x-x';\alpha)=G^{(ab)}_0(x-x';\alpha)-G^{(ab)}_0(x-x')\label{1.16}.
\eea

In the next section, eq. (\ref{1.15}) is used to investigate some finite temperature applications with different choices for the constant vector $u^\mu$ which leads to a violation of the Lorentz symmetry.

\section{Some Applications: TFD and Lorentz violation}

Here, applications for different choices of the compactification parameter $\alpha$ and different directions of the constant vector $u^\mu$ are considered.

\subsection{Stefan-Boltzmann law for massive scalar field}

As a first application, the topology in which one spacetime dimension is compactified is considered, i.e. $\alpha=(\beta,0,0,0)$. Here, the time-axis is compactified into a circumference $\beta$. In this case, the Bogoliubov transformation Eq. (\ref{BT}) and the Green function Eq. (\ref{1.16}) are given, respectively, as
\bea
v^2(\beta)&=&\sum_{l_0=1}^{\infty}e^{-\beta k^0 l_0}\label{2.1},\\
\overline{G}_0(x-x',\beta)&=&2\sum_{l_0=1}^{\infty}G_0(x-x'-i\beta l_0 n_0)\label{2.2},
\eea
where $n_0^\mu=(1,0,0,0)$.

Now, let us analyze the energy-momentum tensor for different choices of the constant vector $u^\mu$.

\subsubsection{Time-like constant vector}

In this case, the constant vector is chosen as $u^\mu=(1,0,0,0)$. For $\mu=\nu=0$, Eq.\,\eqref{1.15} becomes
\bea
{\cal T}^{00(11)}(\beta)=i\lim_{x'\to x}\sum_{l_0=1}^{\infty}{\bigg \{}(1+\lambda)\partial^0\partial'^0+\partial^1\partial'^1+\partial^2\partial'^2+\partial^3\partial'^3-m^2{\bigg \}}G_0(x-x'-i\beta l_0 n_0)\label{2.3}.
\eea
Performing the derivatives and some simplifications, this equation leads to
\bea
{\cal T}^{00(11)}(\beta)=\frac{3m^2}{4\pi^2\beta^2}\sum_{l_0=1}^{\infty}\frac{1}{l_0}\left[(2+\lambda)K_2(m\beta l_0)+\frac{4\lambda m\beta l_0}{3}K_1(m\beta l_0)\right]\label{2.4}.
\eea
This expression is the Stefan-Boltzmann law associated with the massive scalar field with corrections due to the aether term that breaks the Lorentz symmetry.
Let us consider the small mass limit, i.e. $(m\to 0)$. In this limit, the Bessel function is given by
\begin{align}
K_z(\nu)\approx\Gamma(\nu)2^{\nu-1}z^{-\nu}\label{2.5}.
\end{align}
Then Eq.\,\eqref{2.4} reads
\begin{align}
{\cal T}^{00(11)}(\beta)=\dfrac{\pi^2(2+\lambda)}{60\beta^4}\label{2.6}.
\end{align}
This is the Stefan-Boltzmann law for the massless scalar field. It is observed that the parameter, that violates the Lorentz symmetry, contributes to increase the energy density.

\subsubsection{Space-like constant vector}

Here, the following cases are investigated: (i) $u^\mu=(0,1,0,0)$, (ii) $u^\mu=(0,0,1,0)$ and (iii) $u^\mu=(0,0,0,1)$. First, let us consider $u^\mu=(0,1,0,0)$. For this case, the energy-momentum tensor, Eq.\,\eqref{1.15}, becomes
\bea
{\cal T}^{00(11)}(\beta)=i\lim_{x'\to x}\sum_{l_0=1}^{\infty}{\bigg \{}\partial^0\partial'^0+(1-\lambda)\partial^1\partial'^1+\partial^2\partial'^2+\partial^3\partial'^3-m^2{\bigg \}}G_0(x-x'-i\beta l_0 n_0)\label{2.7},
\eea
Taking the component $\mu=\nu=0$, it turns out that
\begin{align}
{\cal T}^{00(11)}(\beta)=\sum_{l_0=1}^{\infty}\frac{1}{l_0^2}\left[\frac{m^2(6-\lambda)K_2(m\beta l_0)}{4\pi^2\beta^2}\right]\label{2.8}.
\end{align}
This is the Stefan-Boltzmann law for a space-like constant vector $u^\mu$. If the small mass limit is taken, this expression becomes
\begin{align}
	{\cal T}^{00(11)}(\beta)=\frac{\pi^2(6-\lambda)}{180\beta^4}\label{2.9}.
\end{align}
It is observed that the other choices for the constant vector, which are $u^\mu=(0,0,1,0)$ and $u^\mu=(0,0,0,1)$ lead to the same results, i.e. Eq. (\ref{2.8}) and Eq. (\ref{2.9}). Furthermore, it is interesting to note that the violation of Lorentz symmetry changes the Stefan-Boltzmann law. However, this modification depends on the direction of the constant vector $u^\mu$. In other words, a time-like constant vector contributes in a different way compared to the contribution due to a space-like constant vector.

\subsection{Casimir effect for the massive scalar field at zero temperature}

In order to calculate the Casimir effect at zero temperature, the compactification parameter is chosen as $\alpha=(0,0,0,i2d)$. With such a choice, the described topology consists of a compactification along the $z$ coordinate, i.e. the topological theory is defined as $\Gamma^1_4=\mathbb{S}^1\times \mathbb{R}^3$ with $\mathbb{S}^1$ being a circumference of length $L=2d$, where $d$ is the distance between the two conducting plates. This spatial compactification leads to the Bogoliubov transformation
\bea
v^2(d)=\sum_{l_3=1}^{\infty}e^{-i2dk^3l_3}\label{2.15},
\eea
and the Green function
\bea
\overline{G}_0(x-x';d)=2\sum_{l_3=1}^{\infty}G_0(x-x'-2dl_3n_3)\label{2.16}
\eea
with $n_3^\mu=(0,0,0,1)$. With these ingredients and for different choices of the constant vector $u^\mu$, the Casimir effect at zero temperature is calculated.

\subsubsection{Time-like constant vector}

Assuming that $u^\mu=(1,0,0,0)$ and $\mu=\nu=0$, the energy-momentum tensor reads
\bea
{\cal T}^{00(11)}(d)=i\lim_{x'\to x}\sum_{l_3=1}^{\infty}{\bigg\{}(1+\lambda)\partial^0\partial'^0+\partial^1\partial'^1+\partial^2\partial'^2+\partial^3\partial'^3-m^2{\bigg\}}G_0(x-x'-2dl_3n_3)\label{2.17},
\eea
that provides us
\begin{align}
{\cal T}^{00(11)}(d)=-\frac{m^2}{8\pi^2d^2}\sum_{l_3=1}^{\infty}\frac{1}{l_3^2}\left[\frac{(2+\lambda)}{2}K_2(2mdl_3)+2mdl_3K_1(2mdl_3)\right]\label{2.18}.
\end{align}
This is the Casimir energy for the massive scalar field with Lorentz symmetry violation. Similarly, the Casimir pressure is obtained for $\mu=\nu=3$. Then
\bea
{\cal T}^{33(11)}(d)=i\lim_{x'\to x}\sum_{l_3=1}^{\infty}{\bigg\{}(1+\lambda)\partial^0\partial'^0-\partial^1\partial'^1-\partial^2\partial'^2+\partial^3\partial'^3+m^2{\bigg\}}G_0(x-x'-2dl_3n_3)\label{2.19}.
\eea
This equation leads to
\begin{align}
{\cal T}^{33(11)}(d)=-\sum_{l_3=1}^{\infty}\frac{1}{l_3^2}\left[\frac{m^2(6+\lambda)K_2(2mdl_3)}{16d^2\pi^2}\right]\label{2.20}.
\end{align}
In the small mass limit, i.e. $ m\to 0 $, the Casimir energy and pressure at zero temperature become
\bea
{\cal T}^{00(11)}(d)&=&-\frac{\pi^2(2+\lambda)}{2880d^4}\label{2.21},\\
{\cal T}^{33(11)}(d)&=&-\frac{\pi^2(6+\lambda)}{2880d^4}\label{2.22}.
\eea
These expressions show that, although the aether term that violates Lorentz symmetry changes the Casimir effect, it does not modify the attractive nature of this phenomenon. Instead, even being very small, it helps to increase the Casimir effect associated with a massive scalar field.

\subsubsection{Space-like constant vector}

First, let us assume $u^\mu=(0,1,0,0)$. Then the component $\mu=\nu=0$ of the energy-momentum tensor reads
\bea
{\cal T}^{00(11)}(d)=i\lim_{x'\to x}\sum_{l_3=1}^{\infty}{\bigg\{}\partial^0\partial'^0+(1-\lambda)\partial^1\partial'^1+\partial^2\partial'^2+\partial^3\partial'^3-m^2{\bigg\}}G_0(x-x'-2dl_3n_3)\label{2.23}.
\eea
This equation gives us the Casimir energy which is given as
\begin{align}
{\cal T}^{00(11)}(d)=-\frac{m^2}{8d^2\pi^2}\sum_{l_3=1}^{\infty}\frac{1}{l_3^2}\left[\frac{(2+\lambda)}{2}K_2(2mdl_3)+2mdl_3K_1(2mdl_3)\right]\label{2.24}.
\end{align}
And for $\mu=\nu=3$ we get
\bea
{\cal T}^{33(11)}(d)=i\lim_{x'\to x}\sum_{l_3=1}^{\infty}{\bigg\{}\partial^0\partial'^0- (1-\lambda)\partial^1\partial'^1-\partial^2\partial'^2+\partial^3\partial'^3+m^2G_0{\bigg\}}G_0(x-x'-2dl_3n_3)\label{2.25},
\eea
which results in the Casimir pressure, i.e.
\begin{align}
{\cal T}^{33(11)}(d)=-\sum_{l_3=1}^{\infty}\frac{1}{l_3^2}\left[\frac{m^2(6-\lambda)K_2(2mdl_3)}{16\pi^2d^2}\right].\label{2.26}
\end{align}
Taking the limit $m\to 0$, Eq. (\ref{2.24}) and Eq. (\ref{2.26}) are given as
\bea
{\cal T}^{00(11)}(d)&=&-\frac{\pi^2(2+\lambda)}{2880d^4}\label{2.27},\\
{\cal T}^{33(11)}(d)&=&-\frac{\pi^2(6-\lambda)}{2880d^4}\label{2.28}.
\eea
In this case, the Lorentz-violating parameter acts against the usual Casimir pressure, i.e., the Casimir force provided by the Lorentz symmetry violation is repulsive. The same results are obtained for the case $u^\mu=(0,0,1,0)$.

Now, let's look at the constant vector $u^\mu=(0,0,0,1)$. In this case, the components $\mu=\nu=0$ and $\mu=\nu=3$ of the energy-momentum tensor are, respectively
\bea
{\cal T}^{00(11)}(d)&=&i\lim_{x'\to x}\sum_{l_3=1}^{\infty}{\bigg\{}\partial^0\partial'^0+\partial^1\partial'^1+\partial^2\partial'^2+(1-\lambda)\partial^3\partial'^3-m^2{\bigg\}}G_0(x-x'-2dl_3n_3)\label{2.29},\\
{\cal T}^{33(11)}(d)&=&i\lim_{x'\to x}\sum_{l_3=1}^{\infty}{\bigg\{}\partial^0\partial'^0-\partial^1\partial'^1-\partial^2\partial'^2+(1-\lambda)\partial^3\partial'^3+m^2{\bigg\}}G_0(x-x'-2dl_3n_3).\label{2.31}
\eea
Performing some calculations, the Casimir energy and pressure are given, respectively, as
\bea
{\cal T}^{00(11)}(d)&=&-\frac{m^2}{8\pi^2d^2}\sum_{l_3=1}^{\infty}\frac{1}{l_3^2}\left[\frac{(2-3\lambda)}{2}K_2(2mdl_3)+(2-\lambda)mdl_3K_1(2mdl_3)\right]\label{2.30},\\
{\cal T}^{33(11)}(d)&=&-\frac{m^2}{16\pi^2d^2}\sum_{l_3=1}^{\infty}\frac{1}{l^2_3}\Bigl[(6-3\lambda)K_2(2mdl_3)-2\lambda mdl_3K_1(2mdl_3)\Bigl]\label{2.32}.
\eea
Following the same steps as in the previous case, the results obtained for the case $m\to 0$ are
\begin{align}
{\cal T}^{00(11)}(d)&=-\frac{\pi^2(2-3\lambda)}{2880d^4}\label{2.33},\\
{\cal T}^{33(11)}(d)&=-\frac{\pi^2(2-\lambda)}{960d^4}\label{2.34}.
\end{align}
Therefore, the corrections for the Casimir effect at zero temperature due to Lorentz symmetry violation depend on the direction of the constant vector $u^\mu$.

\subsection{Casimir effect for the massive scalar field at finite temperature}

In this section, the main point of our discussion is calculated, i.e. the Casimir effect for the massive scalar field at finite temperature in the presence of the aether term that introduces violation of Lorentz symmetry. To achieve this goal, two compactifications are needed. Then the compactification parameter is chosen as $\alpha=(\beta,0,0,i2d)$. For this case, the Bogoliubov transformation is given as
\bea
v^2(\beta,d)=\sum_{l_0=1}^{\infty}e^{\beta k^0l_0}+\sum_{l_3=1}^{\infty}e^{-i2dk^3l_3}+2\sum_{l_0,l_3=1}^{\infty}e^{-\beta k^0l_0-i2dk^3l_3}.\label{2.35}
\eea
The first two terms correspond to the previous cases, while the third term represents the effect of two compactifications together. The Green function for the last term is 
\bea
\overline{G}(x-x';\beta,d)=4\sum_{l_0,l_3=1}^{\infty}G_0(x-x'-i\beta l_0n_0-2dl_3n_3).\label{2.36}
\eea

Now, let us analyze the energy-momentum tensor considering the constant vector $u^\mu$ as a time-like vector in the first analysis and in the sequence as a space-like vector.

\subsubsection{Time-like constant vector}

Considering $u^\mu=(1,0,0,0)$, the third term of Eq. (\ref{2.35}) and Eq. (\ref{2.36}), the energy-momentum tensor with $\mu=\nu=0$ becomes
\bea
{\cal T}^{00(11)}(\beta,d)&=&2i\lim_{x'\to x}\sum_{l_0,l_3=1}^{\infty}{\bigg\{}(1+\lambda)\partial^0\partial'^0+\partial^1\partial'^1+\partial^2\partial'^2+\partial^3\partial'^3-m^2{\bigg\}}\nonumber\\
&\times&G_0(x-x'-i\beta l_0n_0-2dl_3n_3).\label{2.37}
\eea
From this equation, the Casimir energy at finite temperature is obtained as
\bea
{\cal T}^{00(11)}(\beta,d)&=&-\frac{m^2}{\pi^2}\sum_{l_0,l_3=1}^{\infty}{\Bigg\{}\frac{\left[(2dl_3)^2-3(\beta l_0)^2\right]}{\left[(2dl_3)^2+(\beta l_0)^2\right]^2}K_2(m\sqrt{(2dl_3)^2+(\beta l_0)^2})\left(\frac{2+\lambda}{2}\right)\nonumber\\
&+&\left(\frac{2(2dl_3)^2+\lambda(\beta l_0)^2}{2}\right)\frac{mK_1(m\sqrt{(2dl_3)^2+(\beta l_0)^2})}{\left[(2dl_3)^2+(\beta l_0)^2\right]^{3/2}}{\Bigg\}}\label{2.38}.
\eea
This result shows that the Lorentz violation alters the phenomenon in both situations, at zero and finite temperature.

In order to get a more detailed analysis, some interesting limits are considered. As a first attempt, let us take just the first term in the sum of  $l_3$ in Eq.\;\eqref{2.38}. Then
\begin{align}
{\cal T}^{00(11)}(\beta,d)=&-\frac{m^2}{\pi^2}\sum_{l_0=1}^{\infty}{\Bigg\{}\frac{\left(4-3\beta^2l_0^2/d^2\right)}{d^2\left(4+\beta^2l_0^2/d^2\right)}K_2(md\sqrt{4+\beta^2l_0^2/d^2})\left(\frac{2+\lambda}{2}\right)\nonumber\\&+\left(8+\frac{\lambda\beta^2l_0^2}{d^2}\right)\frac{mK_1(md\sqrt{4+\beta^2l_0^2/d^2})}{2d\left(4+\beta^2l_0^2/d^2\right)^{3/2}}{\Bigg\}}\label{2.39}.
\end{align}
Taking the large mass limit ($md \gg 1$), the asymptotic expression for the modified Bessel function is given as
\begin{align}
K_\nu(z)=\sqrt{\frac{\pi}{2z}}e^{-z}\label{2.40}.
\end{align}
Defining $\alpha_{l_0}=\frac{\beta l_0}{d}$, Eq.\;\eqref{2.39} is written as
\begin{align}
{\cal T}^{00(11)}(\beta,d)=\frac{m^{3/2}}{\sqrt{2}\pi^{3/2}d^{3/2}}\sum_{l_0=1}^{\infty}{\bigg\{}\frac{(4-3\alpha^2_{l_0})(2+\lambda)+md(8+\lambda\alpha^2_{l_0})(4+\alpha^2_{l_0})^{1/2}}{2d(4+\alpha^2_{l_0})^{9/4}}{\bigg\}}e^{-md\sqrt{4+\alpha^2_{l_0}}}\label{2.43}.
\end{align}
For low temperatures, i.e. $\beta/d \gg 1$, the last equation becomes
\begin{align}
{\cal T}^{00(11)}(\beta,d)=-\frac{\sqrt{m^5}\lambda}{2\sqrt{2\pi^3}}\frac{1}{\beta^{3/2}}\sum_{l_0=1}^{\infty}\frac{e^{-m\beta l_0}}{l_0^{3/2}}\label{2.44}.
\end{align}
Therefore, at this limit the Casimir energy decays exponentially with $\beta$.

Now the massless case, i.e. $(md=0)$,  is investigated. Using the asymptotic expression for the modified Bessel function Eq. (\ref{2.5}), Eq. (\ref{2.38}) is written in the form
\begin{align}
{\cal T}^{00(11)}(\beta,d)=-\frac{(2+\lambda)}{\pi^2d^4}\sum_{l_0,l_3=1}^{\infty}\frac{(4l_3^2-3\alpha^2_{l_0})}{(4l_3^2+\alpha^2_{l_0})}\label{2.50}.
\end{align}
Performing the sum in $l_3$, it gives
\begin{align}
\sum_{l_3=1}^{\infty}\frac{(4l_3^2-3\alpha^2_{l_0})}{(4l_3^2+\alpha^2_{l_0})^3}&=\frac{1}{16\alpha^4_{l_0}}{\bigg[}24-4\pi\alpha_{l_0}\coth\left(\frac{\pi\alpha_{l_0}}{2}\right)-2\pi^2\alpha^2_{l_0}\mbox{csch}^2\left(\frac{\pi\alpha_{l_0}}{2}\right)\nonumber\\&-\pi^3\alpha^3_{l_0}\coth\left(\frac{\pi\alpha_{l_0}}{2}\right)\mbox{csch}^2\left(\frac{\pi\alpha_{l_0}}{2}\right){\bigg]}\label{2.51}.
\end{align}
In the low temperatures limit, the hyperbolic functions assume the values $ \mbox{csch}\left(\frac{\pi\alpha_{l_0}}{2}\right)\to 0 $ and $ \mbox{coth}\left(\frac{\pi\alpha_{l_0}}{2}\right)\to 1 $. Then
\begin{align}
{\cal T}^{00(11)}(\beta,d)=-\frac{(2+\lambda)}{16\pi^2d^4}\sum_{l_0=1}^{\infty}\frac{1}{\alpha^4_{l_0}}\left[24-4\pi\alpha_{l_0}\right]\label{2.53}.
\end{align}
carrying out the sum, the Casimir energy at such limit reads
\begin{align}
{\cal T}^{00(11)}(\beta,d)=-\frac{(2+\lambda)}{60\pi\beta^4}\left[\pi^3-15\frac{\beta\zeta(3)}{d}\right]\label{2.55},
\end{align}
where $\zeta(3)$ is the Riemann zeta function.

The general expression for the Casimir pressure is given as
\begin{align}
{\cal T}^{33(11)}(\beta,d)=&-\frac{m^2}{\pi^2}\sum_{l_0,l_3=1}^{\infty}{\Bigg\{}\frac{[3(2dl_3)^2-(\beta l_0)^2]}{[(2dl_3)^2+(\beta l_0)^2]}K_2(m\sqrt{(2dl_3)^2+(\beta l_0)^2})\nonumber\\&+\frac{[(2dl_3)^2-3(\beta l_0)^2]}{2[(2dl_3)^2+(\beta l_0)^2]^2}\lambda K_2(m\sqrt{(2dl_3)^2+(\beta l_0)^2)}\nonumber\\&-\frac{(\beta l_0)^2(2+\lambda)}{2[(2dl_3)^2+(\beta l_0)^2]^{3/2}}mK_1(m\sqrt{(2dl_3)^2+(\beta l_0)^2}){\Bigg\}}\label{2.56}.
\end{align}
Similar limits can be analyzed for this result. The main result is that the aether term modifies the Casimir effect at all limits, large mass, no mass, or low temperatures.

\subsubsection{Space-like constant vector}

Assuming that $u^\mu=(0,1,0,0)$ is a space-like vector, the Casimir energy at finite temperature with contribution due to the aether term is
\begin{align}
{\cal T}^{00(11)}(\beta,d)=&-\frac{m^2}{\pi^2}\sum_{l_0,l_3=1}^{\infty}{\Bigg\{}\frac{[(2dl_3)^2-3(\beta l_0)^2]}{[(2dl_3)^2+(\beta l_0)^2]}K_2(m\sqrt{(2dl_3)^2+(\beta l_0)^2})\nonumber\\&+\frac{\lambda K_2(m\sqrt{(2dl_3)^2+(\beta l_0)^2})}{2[(2dl_3)^2+(\beta l_0)^2]}+\frac{m(2dl_3)^2K_1(m\sqrt{(2dl_3)^2+(\beta l_0)^2})}{2[(2dl_3)^2+(\beta l_0)^2]^{3/2}}{\Bigg\}}\label{2.58}.
\end{align}

In order to compare the results, the same limits discussed in the previous subsection are investigated for this case. First, let us consider the large mass limit $md\gg 1$. Then Eq. (\ref{2.58}) becomes
\begin{align}
{\cal T}^{00(11)}(\beta,d)=\frac{m^{3/2}}{\sqrt{2}\pi^{3/2}d^{3/2}}\sum_{l_0=1}^{\infty}{\Bigg\{}\frac{8-6\alpha_{l_0}^2+\lambda(4+\alpha_{l_0}^2)+8md(4+\alpha_{l_0}^2)^{1/2}}{2d^2(4+\alpha_{l_0}^2)^{9/4}}{\Bigg\}}e^{-md\sqrt{4+\alpha_{l_0}^2}}\label{2.59}.
\end{align}
At the limit of low temperatures, we get
\begin{align}
{\cal T}^{00(11)}(\beta,d)=-\frac{\sqrt{m^3}e^{-m\beta}}{2\sqrt{2\pi^3}\beta^{5/2}}(\lambda-6)\label{2.60}.
\end{align}
And for the massless and low temperatures case, the Casimir energy is given as
\begin{align}
{\cal T}^{00(11)}(\beta,d)=-\frac{1}{30\pi\beta^4}\left[\frac{\pi^3(\lambda-6)}{6}+\frac{15\beta(\lambda-4)}{4d}\zeta(3)\right]\label{2.62}.
\end{align}
The Casimir pressure at finite temperature is obtained as
\begin{align}
	{\cal T}^{33(11)}(\beta,d)=&-\frac{m^2}{\pi^2}\sum_{l_0,l_3=1}^{\infty}{\Bigg\{}\frac{[3(2dl_3)^2-(\beta l_0)^2]}{[(2dl_3)^2+(\beta l_0)^2]^2}K_2(m\sqrt{(2dl_3)^2+(\beta l_0)^2})\nonumber\\&-\frac{\lambda K_2(m\sqrt{(2dl_3)^2+(\beta l_0)^2})}{2((2dl_3)^2+(\beta l_0)^2)}-\frac{m\lambda(\beta l_0)^2K_1(m\sqrt{(2dl_3)^2+(\beta l_0)^2})}{[(2dl_3)^2+(\beta l_0)^2]^{3/2}}{\Bigg\}}\label{2.63}.
\end{align}
Again, all previously applied limits can be used for this pressure. The same results are obtained for the space-like vector $u^\mu=(0,0,1,0)$.

As a last case, let us consider $u^\mu=(0,0,0,1)$. Performing similar calculations, the Casimir energy and pressure are given, respectively, as
\begin{align}
	{\cal T}^{00(11)}(\beta,d)=&-\frac{m^2}{\pi^2}\sum_{l_0,l_3=1}^{\infty}{\Bigg\{}\frac{[(2dl_3)^2-3(\beta l_0)^2]}{[(2dl_3)^2+(\beta l_0)^2]^2}K_2(m\sqrt{(2dl_3)^2+(\beta l_0)^2})\nonumber\\&-\frac{[3(2dl_3)^2-(\beta l_0)^2]}{2[(2dl_3)^2+(\beta l_0)^2]^2}\lambda K_2(m\sqrt{(2dl_3)^2+(\beta l_0)^2})\nonumber\\&+\frac{2m(dl_3)^2(2-\lambda)}{[(2dl_3)^2+(\beta l_0)^2]^{3/2}}K_1(m\sqrt{(2dl_3)^2+(\beta l_0)^2}){\Bigg\}}\label{2.65}
\end{align}
and 
\bea
	{\cal T}^{33(11)}(\beta,d)=&-\frac{m^2}{\pi^2}\sum_{l_0,l_3=1}^{\infty}{\Bigg\{}\frac{[3(2dl_3)^2-(\beta l_0)]}{2[(2dl_3)^2+(\beta l_0)^2]^3}K_2(m\sqrt{(2dl_3)^2+(\beta l_0)^2})(2-\lambda)\nonumber\\&-\frac{[2\lambda(dl_3)^2+(\beta l_0)^2]}{[(2dl_3)^2+(\beta l_0)^2]^{3/2}}mK_1(m\sqrt{(2dl_3)^2+(\beta l_0)^2}){\Bigg\}}\label{2.70}.
\eea
In order to avoid repetitions, limits are not applied as the analyzes are very similar to the previous cases. The main observation that must be considered is the fact that the modifications caused by the aether term depend on the direction of the constant vector which is coupled with the derivative of the scalar field.

\section{Conclusions}

The TFD formalism, which is a real-time quantum field theory at finite temperature, allows us to study some applications by choosing different topologies. Here, three different cases are considered. The first case consists of the investigation of effects resulting from the compactification of the time axis. The second case uses a similar topology, however, the compactification takes place along the $z$ coordinate. In the latter case, both compactifications are considered at the same time. Based on this topological theory, different phenomena such as the Stefan-Boltzmann law and the Casimir effect can be considered on an equal footing. In order to analyze these effects, a massive scalar field in a Lorentz-violating theory is introduced. The term that introduces the violation of Lorentz symmetry is the CPT-even aether-like term. Then our results show how the Lorentz symmetry violation modifies the energy density associated with the scalar field and the Casimir effect at zero and non-zero temperature. The aether-like term, which breaks Lorentz, is constructed as the coupling between the derivative of the field and a constant vector $u^\mu$. It is verified that the modifications due to this Lorentz-violating term depend on whether the constant vector is a time-like or a space-like vector. Therefore, the aether term changes both, the Stefan-Boltzmann law and the Casimir effect at zero and finite temperature. In addition, the limits of large mass, small mass, and low temperature are discussed. Although some results discussed here have been calculated in other works, similar analyzes using the TFD formalism have never been done in the literature. In this way, the results obtained are direct and exact, without any simplification or approximation. Within this formalism, the Casimir effect is calculated at zero and finite temperature just by choosing a different topology, and then the same proceeding is considered. It is a unique feature of the TFD formalism. It is important to note that our results at finite temperature, i.e. Eqs. (\ref{2.38}), (\ref{2.56}), (\ref{2.58}), (\ref{2.63}), (\ref{2.65}) and (\ref{2.70}), are consistent with the results obtained in \cite{Cruz2} where the Matsubara approach has been used. Furthermore, although the study developed here is theoretical, if an experimental apparatus is possible,  the effects of temperature can help improve constraints on the Lorentz-violating parameter.

\section*{Acknowledgments}

This work by A. F. S. is supported by National Council for Scientific and Technological Develo\-pment - CNPq projects 430194/2018-8 and 313400/2020-2. L. H. A. R. Ferreira thanks CAPES for financial support.

\end{document}